\documentstyle [11pt, epsfig]{article}
\title{\bf Yukawa Corrections to $\gamma \gamma \rightarrow b\bar{b}$ in the Topcolor Assisted Technicolor
Models }
\author{Jinshu Huang $^{1,2,}$\thanks{Electronic address:
jshuang@vip.sina.com}  \ \ and
Gongru Lu $^{1,}$\thanks{Electronic address: lugongru@sina.com} \\
\small $^1$ Department of Physics, Henan Normal University, Xinxiang
453007, P. R. China \\
\small $^2$ Department of Physics, Nanyang Normal University,
Nanyang
473061, P. R. China \\
}

\begin{document}
\maketitle

\begin{abstract}
We study the Yukawa corrections to the $\gamma\gamma \rightarrow
b\bar{b}$ corss section in the topcolor assisted technicolor models
at the photon-photon colliders. We find that, for the favorable
parameters, the relative corrections from pseudo Goldstone bosons
give out a $3.2\% \sim 5.9\%$ decrement of the cross section from
the tree level when $\sqrt{s}=500\ {\rm GeV}$, the contributions
from new extended technicolor gauge bosons $Z^*$ and colored gauge
bosons $B$ are negligibly small, and the relative correction arising
from new color-singlet heavy gauge boson $Z'$ is less than $-3.2\%$.
Therefore, the total relative corrections are significantly larger
than the corresponding corrections in the standard model, the
general two Higgs doublet model and the minimal supersymmetric
standard model. Since these corrections are obvious for the
International Linear Colliders, the process $\gamma\gamma
\rightarrow b\bar{b}$ is really interesting in testing the standard
model and searching for the signs of technicolor.
\end{abstract}


\section*{I. INTRODUCTION}

\ \ \ The collisions of high energy photons produced at the linear
collider provide a comprehensive laboratory for testing the standard
model (SM) and probing new physics beyond the SM \cite{Brodsky1995}.
With the advent of the new collider technique \cite{Ginzburg1981},
one can obtain the high energy and high intensity photon beams by
using Compton laser photons scattering off the colliding electron
and positron beams, and a large number of heavy quark pairs can be
produced by this method. The photon energy spectrums show that there
are many relatively soft photons, the production of heavy top quark
will be suppressed for reduced collision energies, but no such
suppression effects the relatively light bottom quark
\cite{Halzen1992}. Therefore it is worthy to investigate the
production of the bottom quark pairs in the photon-photon
collisions.

In the SM, this process has been calculated and the QCD threshold
effects of the process also have been examined \cite{Eboli1993}.
Reference \cite{Han1996} has investigated the Yukawa corrections to
this process in both the genernal two Higgs doublet model (2HDM) and
the minimal supersymmetric standard model (MSSM), which shows the
relative corrections to the total cross section of the process
$e^+e^- \rightarrow \gamma\gamma \rightarrow b\bar{b}$ are less than
$0.1\%$ for favorable parameter values. In the paper, we present the
calculation of the Yukawa corrections to this process in the
topcolor assisted technicolor models, which arise from the virtual
effects of the third generation quarks, charged pseudo Goldstone
bosons (PGBs), and new gauge bosons in photon-photon collisions. It
is organized as follows. In Sec. II, we present a brief review of
the original topcolor assisted technicolor (TOPCTC) model and the
multiscale walking topcolor assisted technicolor (TOPCMTC) model. In
Sec. III, we give out the analytical results in term of the
well-known standard notation of one-loop Feynman integrals. The
numerical results and conclusions are included in Sec. IV, and the
form factors appeared in the cross section are presented in the
Appendices A and B.

\section*{II. GENERAL CHARACTERISTICS OF THE TOPCOLOR ASSISTED TECHNICOLOR MODELS}

\ \ \ As we know, technicolor---a strong interaction of fermions and
gauge bosons at the scale $\Lambda_{\rm TC} \sim 1\ {\rm TeV}$--- is
a scenario for the dynamical breakdown of electroweak symmetry to
electromagnetism \cite{Weinberg1979}. Based on the similar
phenomenon of chiral symmetry breakdown in QCD, technicolor is
explicitly defined and completely natural. To account for the mass
of quarks, leptons, and Goldstone ``technipions'' in such a scheme,
technicolor, ordinary color, and flavor symmetry are embedded in a
large gauge group, called extended technicolor (ETC)
\cite{Dimopoulos1979}. Because of the conflict between constraints
on flavor-changing neutral currents and the magnitude of
ETC-generated quark, lepton and technipion masses, classical
technicolor was superseded by a ``walking'' technicolor and
``multiscale technicolor'' \cite{Holdom1981,Lane1989}. The
incapability of explain the top quark's large mass without a clash
of either cherished  notions of naturalness or experiments from the
$\rho$ parameter and the ${\em Z}\rightarrow b\bar{b}$ decay rate by
ETC \cite{Chivukula1992} led to the original topcolor assisted
technicolor by C. T. Hill \cite{Hill1995} and the multiscale walking
topcolor assisted technilor model \cite{Lane1995}.

The original TOPCTC model assumes \cite{Hill1995,Hill1991,Lu1997}:
(i) electroweak interactions are broken by technicolor; (ii) the top
quark mass is large because it is the combination of a dynamical
condensate component $(1-\varepsilon)m_t$, generated by a new strong
dynamics, together with a small fundamental component $\varepsilon
m_t (\varepsilon \sim 0.03-0.1)$, generated by ETC; (iii) the new
strong dynamics is assumed to be chiral critically strong but
spontaneously broken by technicolor at the scale $\sim 1~{\rm TeV}$,
and it generally couples preferentially to the third  generation.
This needs a new class of technicolor models incorporating
``topcolor'' (TOPC). The dynamics at $\sim 1~{\rm TeV}$ scale
involves the gauge structure
\begin{equation}
 SU(3)_1\times SU(3)_2\times U(1)_{Y_1}\times U(1)_{Y_2}
\rightarrow SU(3)_{\rm QCD}\times U(1)_{\rm EM} \nonumber
\end{equation}
where $SU(3)_1\times U(1)_{Y_1}~[SU(3)_2\times U(1)_{Y_2}]$
generally couples preferentially to the third (first and second)
generation, and is assumed to be strong enough to form chiral
$<\bar{t}t>$ but not $<\bar{b}b>$ condensation by the $U(1)_{Y_1}$
coupling.  A residual global symmetry $SU(3)' \times U(1)'$ implies
the existence of a massive color-singlet heavy $Z'$ and an octet
$B$. A symmetry-breaking pattern onlined above will generically give
rise to three top pions, $\pi_{t}$, near the top mass scale.

The couplings of the gauge bosons $Z'$ and $B$ to bottom quark given
by the topcolor interactions which for the process $\gamma\gamma
\rightarrow b\bar{b}$ can be written as
\begin{eqnarray}
Z'b\bar{b}: && \frac{1}{6}g_1 \cot \theta'\gamma^{\mu}L-
\frac{1}{3}g_1 \cot
 \theta'\gamma^{\mu}R, \\
B b\bar{b}: && \frac{1}{2} g_3 \cot\theta \lambda^a \gamma^{\mu},
\end{eqnarray}
where $L,R=(1 \mp \gamma_5)/2$ are the left- and right-handed
projectors, $\lambda^a$ is a Gell-Mann matrix acting on ordinary
color indices, $g3$ ($g_1$) is the QCD $U(1)_Y$ coupling constant at
the scale $\sim 1~{\rm TeV}$. The SM $U(1)_Y$ field $B_{\mu}$ and
the $U(1)'$ field $Z_{\mu}'$ are then defined by orthogonal rotation
with mixing angle $\theta$ ($\theta'$). If we take
\begin{eqnarray}
\kappa=\frac{g_3^2 \cot^2\theta}{4\pi},\ \  \kappa_1=\frac{g_1^2
\cot^2\theta '}{4\pi},
\end{eqnarray}
Ref. \cite{Hill2003} shows that the value of $\kappa$ must be
approximately 2 and $\kappa_1$ is assumed to be $O(1)$.

There exists the ETC gauge bosons $Z^*$ including the sideways and
diagonal gauge bosons in this model. The coupling of $Z^*$ to the
fermions and technifermions can be found in Ref. \cite{Wu1995}. For
the sake of simplicity, we assume that the mass of the sideways
gauge boson is equal to the mass of the diagonal gauge boson, namely
$m_{Z^*}$, so the $Z^* b\bar{b}$ coupling by the ETC dynamics can be
given by
\begin{eqnarray}
 Z^* b\bar{b}: \ \ \ \ -\frac{\varepsilon m_t}{16\pi f_{\pi}} \frac{e}{s_Wc_W}
[\frac{N_C}{N_{TC}+1}\xi_t(\xi_t^{-1}+\xi_b)-\xi_t^2]\gamma_{\mu} L,
\end{eqnarray}
where $N_{TC}$ and $N_C$ are the numbers of technicolors and
ordinary colors, respectively; $s_W = \sin \theta_W$ and $c_W = \cos
\theta_W$ with $\theta_W$ being the Weinberg angle; $\xi_t$ and
$\xi_b$ are coupling coefficients and are ETC gauge-group-dependent.
Following Ref. \cite{Wu1995}, we take $\xi_t=1/\sqrt{2}$ and
$\xi_{b}=0.028 \xi_t^{-1}$.

In this TOPCTC model, there are 60 technipions in the ETC sector
with decay constant $f_{\pi}=123$ GeV and three top pions $\pi_t^0,
\pi_t^{\pm}$ in the TOPC sector with decay constant $f_{\pi_t}=50$
GeV. The ETC sector is one generation technicolor model
\cite{Dimopoulos1979}. The relevant technipions in this study are
only the color-singlet $\pi$ and color-octet $\pi_8$. The
color-singlet (octet) technipion-top (bottom) interactions are given
by
\begin{eqnarray}
&&\frac{c_t \varepsilon m_t}{\sqrt{2}f_{\pi}} [i\bar{t}\gamma_5t\pi^0
+i\bar{t}\gamma_5 t\pi^3+ \frac{1}{\sqrt{2}}\bar{t}(1-\gamma_5)b\pi^+
+\frac{1}{\sqrt{2}} \bar{b}(1+\gamma_5)t\pi^-], \\
&&\frac{\sqrt{2} \varepsilon m_t}{f_{\pi}} [i\bar{t}\gamma_5
\frac{\lambda^a}{2}
t\pi_8^0 +i\bar{t}\gamma_5\frac{\lambda^a}{2}t\pi_8^3+ \frac{1}{\sqrt{2}}
\bar{t}(1-\gamma_5)\frac{\lambda^a}{2}b\pi_8^+ +\frac{1}{\sqrt{2}}\bar{b}
(1+\gamma_5)\frac{\lambda^a}{2}t\pi_8^-],
\end{eqnarray}
with the coefficient $c_t=1/\sqrt{6}$.

The coupling of the top pions to the top (bottom) quark has the form
\begin{equation}
\frac{(1-\varepsilon)m_t}{\sqrt{2}f_{\pi_t}}[i \bar{t}\gamma_5 t \pi_t^0
 +\frac{1}{\sqrt{2}}\bar{t}(1-\gamma_5)b\pi_t^+
 +\frac{1}{\sqrt{2}}\bar{b}(1+\gamma_5)t\pi_t^-].
\end{equation}

The interaction of the gauge boson $\gamma$ and the top pions
$\pi_t^{\pm}$ is
\begin{equation}
ie(p'-p)^{\mu},
\end{equation}
which $p'$, $p$ denote the momentums of $\pi_t^+$ and $\pi_t^-$,
respectively. More detail Feynman rules needed in the calculations
can be found in Refs. \cite{Ellis1981} and \cite{Miransky1989}.

For the topcolor assisted multiscale technicolor (TOPCMTC) model
\cite{Lane1995,Yue1997}, it is different from the original TOPCTC
model mainly by the ETC sector. In the original TOPCTC model, the
ETC sector is the one generation technicolor model with
$f_{\pi}=123$ GeV, $c_t=1/\sqrt{6}$ and $N_{TC}=4$, and in TOPCMTC
model the ETC sector is the multiscale walking technicolor model
with $f_{\pi}=40$ GeV, $c_t=2/\sqrt{6}$ and $N_{TC}=6$
\cite{Lane1995,Yue1997}.

\section*{III. YUKAWA CORRECTIONS TO THE BOTTOM PAIR
PRODUCTION IN PHOTON-PHOTON COLLISIONS}

\ \ \ The relevant Feynman diagrams for the corrections arising from
PGBs to the $\gamma \gamma \rightarrow b\bar{b}$ production
amplitudes are shown in Figs. 1 (c)-(m). In our calculation, we use
the dimensional regularization to regulate all the ultraviolet
divergences in the virtual loop corrections, and adopt the Feynman
gauge and on-mass-shell renormalization scheme \cite{Bohm1986}. The
renormalized amplitude for $\gamma \gamma \rightarrow b\bar{b}$
contains
\begin{equation}
M_{\rm ren}=M_0+ \delta M =M_0+ \delta M^{\rm self} + \delta M^{\rm
vertex}+ \delta M^{\rm box}+ \delta M^{\rm tr},
\end{equation}
where $M_0$ is the amplitude at the tree level, $\delta M^{\rm
self}$, $\delta M^{\rm vertex}$, $\delta M^{\rm box}$ and $\delta
M^{\rm tr}$ represent the Yukawa corrections arising from the
self-energy, vertex, box, and triangle diagrams, respectively. Their
explicit forms are given by
\begin{eqnarray}
M_0=M_0^{\hat{t}}+M_0^{\hat{u}},
\end{eqnarray}
\begin{eqnarray}
\delta M^{\rm self}=\delta M^{s(\hat{t})}+\delta M^{s(\hat{u})},
\end{eqnarray}
\begin{eqnarray}
\delta M^{\rm vertex}=\delta M^{v(\hat{t})}+\delta M^{v(\hat{u})},
\end{eqnarray}
\begin{eqnarray}
\delta M^{\rm box}=\delta M^{b(\hat{t})}+\delta M^{b(\hat{u})},
\end{eqnarray}
where
\begin{eqnarray}
M_0^{\hat{t}}=-i\frac{e^2 Q_b^2}{\hat{t}-m_b^2} \epsilon_{\mu} (p_4)
\epsilon_{\nu}(p_3) \bar{u}(p_2) \gamma^{\mu} (\not p_2 -\not
p_4+m_b) \gamma^{\nu} v(p_1),
\end{eqnarray}
\begin{eqnarray}
M_0^{\hat{u}}=M_0^{\hat{t}}(p_3 \leftrightarrow p_4, \hat{t}
\leftrightarrow \hat{u}),
\end{eqnarray}
\begin{eqnarray}
\delta M^{s(\hat{t})}=i\frac{e^2 Q_b^2}{(\hat{t}-m_b^2)^2}
\epsilon_{\mu} (p_4) \epsilon_{\nu}(p_3) \bar{u}(p_2)
[f_1^{s(\hat{t})} \gamma^{\mu} \gamma^{\nu} + f_2^{s(\hat{t})}
p_2^{\mu} \gamma^{\nu}+f_3^{s(\hat{t})} \not p_4 \gamma^{\mu}
\gamma^{\nu}] v(p_1),
\end{eqnarray}
\begin{eqnarray}
\delta M^{s(\hat{u})}=\delta M^{s(\hat{t})} (p_3 \leftrightarrow
p_4, \hat{t} \leftrightarrow \hat{u}),
\end{eqnarray}
\begin{eqnarray}
\delta M^{v(\hat{t})}&=&-i\frac{e^2 Q_b}{\hat{t}-m_b^2}
\epsilon_{\mu} (p_4) \epsilon_{\nu}(p_3) \bar{u}(p_2)
[f_1^{v(\hat{t})} \gamma^{\mu} \gamma^{\nu} + f_2^{v(\hat{t})}
\gamma^{\mu} p_1^{\nu}+ f_3^{v(\hat{t})} p_2^{\mu} \gamma^{\nu}+
f_4^{v(\hat{t})} p_2^{\mu} p_1^{\nu} \nonumber\\
&&+f_5^{v(\hat{t})} \not p_4 \gamma^{\mu} \gamma^{\nu}
+f_6^{v(\hat{t})} \not p_4 \gamma^{\mu} p_1^{\nu} +f_7^{v(\hat{t})}
\not p_4 p_2^{\mu} \gamma^{\nu}] v(p_1),
\end{eqnarray}
\begin{eqnarray}
\delta M^{v(\hat{u})}=\delta M^{v(\hat{t})} (p_3 \leftrightarrow
p_4, \hat{t} \leftrightarrow \hat{u}),
\end{eqnarray}
\begin{eqnarray}
\delta M^{b(\hat{t})}&=&-i\frac{e^2}{16\pi^2} \epsilon_{\mu} (p_4)
\epsilon_{\nu}(p_3) \bar{u}(p_2) [f_1^{b(\hat{t})} \gamma^{\mu}
\gamma^{\nu} + f_2^{b(\hat{t})} \gamma^{\nu}\gamma^{\mu}  +
f_3^{b(\hat{t})} \gamma^{\mu} p_1^{\nu}+ f_4^{b(\hat{t})} p_1^{\mu}
\gamma^{\nu} + f_5^{b(\hat{t})} \gamma^{\mu} p_2^{\nu}\nonumber\\
&&+ f_6^{b(\hat{t})} p_2^{\mu} \gamma^{\nu} +f_7^{b(\hat{t})}
p_1^{\mu} p_1^{\nu} +f_8^{b(\hat{t})} p_1^{\mu} p_2^{\nu}
+ f_9^{b(\hat{t})} p_2^{\mu} p_1^{\nu} +f_{10}^{b(\hat{t})} p_2^{\mu} p_2^{\nu}
+f_{11}^{b(\hat{t})} \not p_4 \gamma^{\mu} \gamma^{\nu} \nonumber\\
&& +f_{12}^{b(\hat{t})} \not p_4 \gamma^{\nu} \gamma^{\mu}
+f_{13}^{b(\hat{t})} \not p_4 \gamma^{\mu} p_1^{\nu}
+f_{14}^{b(\hat{t})} \not p_4 p_1^{\mu} \gamma^{\nu}
+f_{15}^{b(\hat{t})} \not p_4 \gamma^{\mu} p_2^{\nu}
+f_{16}^{b(\hat{t})} \not p_4 p_2^{\mu} \gamma^{\nu} \nonumber\\
&&+f_{17}^{b(\hat{t})} \not p_4 p_1^{\mu} p_1^{\nu}
+f_{18}^{b(\hat{t})} \not p_4 p_1^{\mu} p_2^{\nu}
+f_{19}^{b(\hat{t})} \not p_4 p_2^{\mu} p_1^{\nu}
+f_{20}^{b(\hat{t})} \not p_4 p_2^{\mu} p_2^{\nu} ] v(p_1),
\end{eqnarray}
\begin{eqnarray}
\delta M^{b(\hat{u})}=\delta M^{b(\hat{t})} (p_3 \leftrightarrow
p_4, \hat{t} \leftrightarrow \hat{u}),
\end{eqnarray}
and
\begin{eqnarray}
\delta M^{\rm tr}=i\frac{e^2}{8\pi^2} f_1^{\rm tr}
g^{\mu\nu}\epsilon_{\mu} (p_4) \epsilon_{\nu}(p_3) \bar{u}(p_2)
 v(p_1).
\end{eqnarray}
Here $\hat{t}=(p_4-p_2)^2$, $\hat{u}=(p_4-p_1)^2$, $p_3$ and $p_4$
denote the momentum of the two incoming photons, and $p_2$ and $p_1$
are the momentum of the outgoing bottom quark and its antiparticle.

The form factors $f_i^{s(\hat{t})}$, $f_i^{v(\hat{t})}$,
$f_i^{b(\hat{t})}$ and $f_i^{\rm tr}$ are expressed in terms of two-
, three-, and four- point scalar integrals, and are presented in the
Appendix A. The basic two-, three-, and four- scalar integrals are
given in Ref. \cite{Clements1983}. It is easy to find that all the
ultraviolet divergences cancel in the effective vertex.

For the new gauge bosons ($Z^*$, $Z'$ and $B$), we plot the relevant
Feynman diagrams for the contributions arising from these particles
to the $\gamma\gamma \rightarrow b\bar{b}$ production amplitudes in
Fig. 2. The form factors from these new gauge bosons are similar to
those of PGBs, and are given in the Appendix B.

The cross section of the subprocess $\gamma\gamma \rightarrow
b\bar{b}$ for the unpolarized photons is given by
\begin{equation}
\hat{\sigma}(\hat{s})=\frac{N_C}{16\pi\hat{s}^2}
\int_{\hat{t}^-}^{\hat{t}^+} {\rm d} \hat{t} \overline{\sum_{\rm
spins}}|M_{\rm ren}(\hat{s},\hat{t})|^2,
\end{equation}
where
\begin{equation}
\hat{t}^{\pm}=(m_b^2-\frac{1}{2}\hat{s}) \pm \frac{1}{2}\hat{s}
\sqrt{1-4 m_b^2/\hat{s}}.
\end{equation}
The bar over the sum recalls averaging over initial spins and
\begin{equation}
\overline{\sum_{\rm spins}}|M_{\rm ren}(\hat{s},\hat{t})|^2=
\overline{\sum_{\rm spins}}|M_{0}|^2+2{\rm Re} \overline{\sum_{\rm
spins}}M_0^{\dagger}\delta M.
\end{equation}

The total cross section $\sigma(s)$ for the bottom pair production
 in $\gamma\gamma$ collisions can be obtained by
folding the elementary cross section $\sigma(\hat{s})$ for the
subprocess $\gamma\gamma \rightarrow b\bar{b}$ with the photon
luminosity at the $e^+e^-$ colliders given in Refs. \cite{Eboli1993}
and \cite{Han1996}, i.e.,
\begin{equation}
\sigma(s)=\int_{2m_b/\sqrt{s}}^{x_{\rm max}} {\rm d}z \frac{{\rm
d}L_{\gamma\gamma}}{{\rm d}z}\hat{\sigma}(\hat{s}) \ \ (\gamma\gamma
\rightarrow b\bar{b} \ {\rm at} \ \hat{s}=z^2 s),
\end{equation}
where $\sqrt{s}$ and $\sqrt{\hat{s}}$ are the $e^+e^-$ and
$\gamma\gamma$ center-of-mass energies respectively, and ${\rm
d}L_{\gamma\gamma}/{\rm d}z$ is the photon luminosity, which can be
expressed as
\begin{equation}
 \frac{{\rm
d}L_{\gamma\gamma}}{{\rm d}z}=2z\int_{z^2/x_{\rm max}}^{x_{\rm max}}
\frac{{\rm d}x}{x} F_{\gamma/e}(x)F_{\gamma/e}(z^2/x),
\end{equation}
For unpolarized initial electron and laser beams, the energy
spectrum of the backscattered photon is given by
\cite{Eboli1993,Cheung1993}
\begin{equation}
 F_{\gamma/e}(x)=\frac{1}{D(\xi)}[1-x+\frac{1}{1-x}
 -\frac{4x}{\xi(1-x)}+\frac{4x^2}{\xi^2
 (1-x^2)}],
\end{equation}
with
\begin{equation}
 D(\xi)=(1-\frac{4}{\xi}-\frac{8}{\xi^2}){\rm ln}(1+\xi) +\frac{1}{2}
 +\frac{8}{\xi}-\frac{1}{2(1+\xi)^2},
\end{equation}
where $\xi=4E_e E_0/m_e^2$ in which $m_e$ and $E_e$ denote the
incident electron mass and energy, respectively, $E_0$ denotes the
initial laser photon energy, and $x=E/E_e$ is the fraction which
represents the ratio between the scattered photon and initial
electron energy for the backscattered photons moving along the
initial electron direction. $F_{\gamma/e}(x)$ vanishes for $x>x_{\rm
max}=E_{\rm max}/E_e=\xi/(1+\xi)$. In order to avoid the creation of
$e^+e^-$ pairs by the interaction of the incident and backscattered
photons, we require $E_0 x_{\rm max} \leq m_e^2/E_e$ which implies
$\xi \leq 2+2\sqrt{2} \approx 4.8$ \cite{Cheung1993}. For the choice
$\xi=4.8$, it can obtain
\begin{equation}
x_{\rm max} \approx 0.83, \ \ D(\xi) \approx 1.8.
\end{equation}

\section*{V. NUMERICAL RESULTS AND CONCLUSIONS}

\subsection*{A. The PGBs Contributions}

\ \ \ It is necessary to point out that, in the calculation of
$\hat{\sigma}(\hat{s})$, instead of calculating the square of the
amplitude $M_{\rm ren}$ analytically, we calculate the amplitudes
numerically by using the method of Ref. \cite{Hagiwara1989}. Care
must be taken in the calculation of the form factors expressed in
terms of the standard loop integrals. As has been discussed in Ref.
\cite{Denner1994}, the formulas for the form factors given in terms
of the tensor loop integrals will be ill defined when the scattering
is forward or backward wherein the Gram determinants of some vanish
and thus their inverses do not exist. This problem can be solved by
taking kinematic cuts on the rapidity $y$ and the transverse
momentum $p_T$. In this paper, we take
\begin{equation}
|y|<2.5, \ \ p_T>20\ {\rm GeV}.
\end{equation}
The cuts will also increase the relative correction
\cite{Beenakker1994}.

In our numerical evaluation, we take a set of independent input
parameters which are known from current experiment. The input
parameters are $m_t=174.2$ GeV, $m_b=4.7$ GeV, $G_F=1.166392 \times
10^{-5}{\rm GeV}^{-2}$, $\sin^2\theta_W=0.2315$, and
$\alpha=1/137.036$ \cite{Yao2006}. It is known that the cross
section for the $e^+e^- \rightarrow \gamma\gamma \rightarrow
b\bar{b}$ at the tree level is model independent, but the quantum
corrections are model dependent. The values of the tree-level cross
section are 7.962 pb, 3.040 pb, and 1.668 pb for $\sqrt{s}=0.5, 1.0$
and $1.5$ TeV, respectively.

Since the ETC sector of this model is one generation technicolor
model. The masses of PGBs are model dependent. In Ref.
\cite{Ellis1981}, the masses of $\pi$ and $\pi_8$ are taken to be in
the range $60\ {\rm GeV}< m_{\pi}< 200\ {\rm GeV}$, $200\ {\rm GeV}<
m_{\pi_8}< \ 500\ {\rm GeV}$. In the TOPC sector, the mass of the
top pion, $m_{\pi_t}$, a reasonable value of the parameter is around
200 GeV. In the following calculation,  we would rather take a
slightly larger range, $150\ {\rm GeV}< m_{\pi_t}< 450\ {\rm GeV}$,
to see its effect, and shall take the masses of $m_{\pi}$, 150 GeV,
and $m_{\pi_8}$, 246 GeV. The final numerical results are plotted in
Figs. 3-5.

Figure 3 shows the relative correction $\delta\sigma(e^+e^-
\rightarrow \gamma\gamma \rightarrow b\bar{b})$ versus $\varepsilon$
with $m_{\pi_t}=225$ GeV, and $\sqrt{s}= 0.5, 1.0, 1.5$ TeV. One can
see that (i) the relative corrections are negative and are between
$-4\%$ and $-7\%$ in general, (ii) the relative corrections decrease
with $\varepsilon$ slowly, which it is natural since the less
$\varepsilon$, the larger contribution can be afforded by the TOPC
sector of this model, (iii) the maximum of the relative corrections
is $-6.8\%$ for $\varepsilon=0.03$, when $\sqrt{s}=1.0$ TeV.

Fig. 4 presents the plots of relative correction
$\delta\sigma(e^+e^- \rightarrow \gamma\gamma \rightarrow b\bar{b})$
vs $m_{\pi_t}$ with $\varepsilon=0.06$, and $\sqrt{s}= 0.5, 1.0,
1.5$ TeV. From this figure, we can see the following: (i) The
relative corrections decrease with $m_{\pi_t}$ sensitively. (ii) The
relative corrections at $\sqrt{s}=1.0$ TeV are larger than those at
$\sqrt{s}=0.5$ TeV and $\sqrt{s}=1.5$ TeV. (iii) The maximum of the
relative corrections can reach $-7.9\%$ for $\varepsilon=0.06$ and
$m_{\pi_t}=150$ GeV when $\sqrt{s}=1.0$ TeV.

Next, we look at the total cross section of the process $e^+e^-
\rightarrow \gamma\gamma \rightarrow b\bar{b}$ arising from PGBs
contribution. We take the case of $m_{\pi}=150$ GeV, $m_{\pi_8}=246$
GeV, $m_{\pi_t}=225$ GeV, and $\varepsilon=0.06$ as an example, and
plot $\sigma(s)$ as a function of $\sqrt{s}$ in Fig. 5. From the
graph, we can find that (i) differing from $\gamma\gamma \rightarrow
t\bar{t}$ \cite{Halzen1992,Cheung1993}, the total cross section of
the process $e^+e^- \rightarrow \gamma\gamma \rightarrow b\bar{b}$
decreases with $\sqrt{s}$ in the range $0.5 \sim 1.5$ TeV. (ii) the
difference between the TOPCTC model and the tree level is smooth,
and has not obvious fluctuation.

For the TOPCMTC model, our calculations show that the contribution
from PGBs in the TOPCMTC model is slightly larger than that of the
original TOPCTC model and the difference is negligibly small.
Therefore the relative corrections $\delta\sigma(e^+e^- \rightarrow
\gamma\gamma \rightarrow b\bar{b})$ and the total cross section
$\sigma(e^+e^- \rightarrow \gamma\gamma \rightarrow b\bar{b})$ in
this model are not plotted one by one.

\subsection*{B. The Gauge Boson Contributions}

\ \ \ Now let us consider the contributions from new gauge bosons to
the $\gamma\gamma \rightarrow b\bar{b}$ cross section.

Firstly, for the ETC gauge boson $Z^*$, we find that, the maximum of
the relative corrections $\delta \sigma _{Z^*}$ is only the order of
$10^{-9} \sim 10^{-10}$ whatever $\varepsilon$, $\sqrt{s}$, and
$m_{Z^*}$ taken in the favorable parameter ranges, and therefore,
can be neglected safely.

Secondly, for the corrections arising from the color-singlet heavy
gauge boson $Z'$, in our calculation we assume the mass of the gauge
boson $Z'$ varying from $300\ {\rm GeV}$ to $1200\ {\rm GeV}$ to
study the effects of $Z'$ \cite{Hill2003}. The numerical results are
plotted in Fig. 6. From this figure, we can find that (i) the
relative corrections are negative and undulate but not as distinctly
as $m_{Z'}$ increases, (ii) when $\kappa_1=1$, $4$ and $8$, the
values of relative correction aren't more than $-0.4\%$, $-1.6\%$
and $-3.2\%$, respectively.

Finally, for the new colored gauge bosons $B$, our calculations
present that the relative correction from these particles is only
the order of $10^{-4} \sim 10^{-5}$ due to their heavy masses, and
is negligibly small.

For the TOPCMTC model, our calculations indicate that the
contribution from $Z^*$ in the TOPCMTC model is slightly larger than
that of the original TOPCTC model but can be still neglected safely,
and the contributions from $Z'$ and $B$ are the same as those of the
original TOPCTC model.

We know, the International Linear Collider (ILC) is the important
next generation linear collider. According to the ILC Reference
Design Report \cite{Brau2007}, the ILC is determined to run with
$\sqrt{s}=500$ GeV and the total luminosity required is $L=500\ {\rm
fb}^{-1}$ with the first four years of operation and $L=1000\ {\rm
fb}^{-1}$ during the first phase of opertion with  $\sqrt{s}=500$
GeV. It means that, millions of the bottom pairs per year can be
produced, and it can also give obvious changes that the $-3.2\% \sim
-5.9\%$ difference of relative corrections are arising from PGBS
contributing in the TOPCTC model when $\sqrt{s}=500$ GeV.
Furthermore, the new gauge boson $Z'$ can also afford a less than
$-3.2\%$ relative correction. But this relative correction is less
than $0.1\%$ in the 2HDM and MSSM \cite{Han1996}, and for the SM,
our calculation shows that this difference from Higgs boson in the
SM is only the order of $10^{-6}$, and is negligibly small.
Therefore via the process $\gamma\gamma \rightarrow b\bar{b}$, the
topcolor assisted technicolor models are experimentally
distinguishable from the SM, 2HDM and MSSM, which affords the
possibility of testing the topcolor assisted technicolor models.

In conclusion, we have calculated the Yukawa corrections to the
process $\gamma\gamma \rightarrow b\bar{b}$ in the topcolor assisted
technicolor models. We find that, for the favorable parameters, the
relative corrections from pseudo Goldstone bosons give out a $3.2\%
\sim 5.9\%$ decrement of the cross section from the tree level when
$\sqrt{s}=500\ {\rm GeV}$, the contributions from new ETC gauge
bosons $Z^*$ and colored gauge bosons $B$ are negligibly small, and
the relative correction arising from new color-singlet heavy gauge
boson $Z'$ is less than $-3.2\%$. Therefore, these corrections are
obvious for the International Linear Colliders and are really
interesting in testing the standard model and searching for the
signs of technicolor.

\section*{ACKNOWLEDGMENTS}

\ \ \ This project was supported by the Natural Science Foundation
of Henan Educational Committee under No. 2007140013, the Natural
Science Foundation of Henan Province under No. 0611050300, and the
National Natural Science Foundation of China under Grant No.
10575029.

\section*{APPENDIX A: THE FORM FACTORS OF PGBS CONTRIBUTION}

\ \ \ The form factors $f_i^{s(\hat{t})}$ of the PGBs contribution
can be expressed by
\begin{eqnarray}
f_1^{s(\hat{t})}=-2m_b(p_2 \cdot p_4)
[\Sigma_S^b(\hat{t})-\frac{\delta m_b}{m_b}-\delta Z_V^b]-2m_b(p_2
\cdot p_4) [\Sigma_V^b(\hat{t})+\delta Z_V^b],\nonumber
\end{eqnarray}
\begin{eqnarray}
f_2^{s(\hat{t})}=4m_b^2 [\Sigma_S^b(\hat{t})-\frac{\delta
m_b}{m_b}-\delta Z_V^b]+4(m_b^2-p_2 \cdot p_4)
[\Sigma_V^b(\hat{t})+\delta Z_V^b],\nonumber
\end{eqnarray}
\begin{eqnarray}
f_3^{s(\hat{t})}=\frac{1}{2}f_2^{s(\hat{t})},\nonumber
\end{eqnarray}
where $\Sigma^b$, $\delta m_b$, and $\delta Z_V^b$ are Yukawa
contribution part of the unrenormalized self-energy function,
$b$-quark mass, and wave function renormalization constants,
respectively. Their expressions are listed as
\begin{eqnarray}
\Sigma^b(p^2)=\not p [\Sigma_V^b(p^2)+\gamma_5 \Sigma_A^b(p^2)]+m_b
\Sigma_S^b(p^2).\nonumber
\end{eqnarray}
Actually the $\Sigma_A^b$ does not contribute to the form factor
$f_i^{s(\hat{t})}$ since the term with $\Sigma_A^b$ includes
$\gamma_5$:
\begin{eqnarray}
\Sigma_V^b(p^2)=-\frac{\lambda_i^2}{32\pi^2}B_1(p^2, m_t,
m_i),\nonumber
\end{eqnarray}
\begin{eqnarray}
\Sigma_A^b(p^2)=\frac{\lambda_i^2}{32\pi^2}B_1(p^2, m_t,
m_i),\nonumber
\end{eqnarray}
\begin{eqnarray}
\Sigma_S^b(p^2)=0,\nonumber
\end{eqnarray}
\begin{eqnarray}
\delta m_b=m_b[\Sigma_V^b(m_b^2)+\Sigma_S^b(m_b^2)],\nonumber
\end{eqnarray}
\begin{eqnarray}
\delta Z_V^b=-\Sigma_V^b(m_b^2)-2m_b^2\frac{\partial}{\partial
p^2}[\Sigma_V^{b}(p^2)+\Sigma_S^{b}(p^2)]|_{p^2=m_b^2}.\nonumber
\end{eqnarray}

The form factors $f_i^{v(\hat{t})}$, $f_i^{b(\hat{t})}$, and
$f_i^{\rm{tr}}$ are given by
\begin{eqnarray}
f_1^{v(\hat{t})}=-\frac{\lambda_i^2 Q_t}{16\pi^2} m_b p_2 \cdot p_4
(C_0^2+C_{11}^2+C_0^4+C_{11}^4),\nonumber
\end{eqnarray}
\begin{eqnarray}
f_2^{v(\hat{t})}=\frac{\lambda_i^2}{8\pi^2} p_2 \cdot p_4[
(C_{12}^3+C_{23}^3)+Q_t (C_{12}^4+C_{23}^4)],\nonumber
\end{eqnarray}
\begin{eqnarray}
f_3^{v(\hat{t})}&=&\frac{\lambda_i^2}{8\pi^2}
[-m_b^2(C_{11}^1+C_{21}^1)+ p_2 \cdot p_4
(C_{12}^1+C_{23}^1)+(-C_{24}^1+C_{24}^3)]\nonumber\\&&+\frac{\lambda_i^2
Q_t}{16\pi^2} [(m_t^2+m_b^2)(C_0^2+C_0^4) +2m_b^2C_{11}^2+2p_2 \cdot
p_4 ( C_{12}^4+C_{23}^4) +m_b^2(C_{21}^2-C_{21}^4)\nonumber\\&&
-2(C_{24}^2+C_{24}^4)]-4Q_t \delta Z_V^b,\nonumber
\end{eqnarray}
\begin{eqnarray}
f_4^{v(\hat{t})}=-\frac{\lambda_i^2}{8\pi^2} m_b
[(C_{11}^3+C_{21}^3)+ Q_t (C_{11}^4+C_{21}^4)],\nonumber
\end{eqnarray}
\begin{eqnarray}
f_5^{v(\hat{t})}&=&\frac{\lambda_i^2}{16\pi^2}
[-C_{24}^1+C_{24}^3]+\frac{\lambda_i^2 Q_t}{32\pi^2}
[(m_t^2+m_b^2)(C_0^2+C_0^4)+2p_2 \cdot p_4
(C_{12}^2+C_{23}^2+C_{12}^4+C_{23}^4) \nonumber \\
&&-m_b^2(C_{21}^2+C_{21}^4)-2(C_{24}^2+C_{24}^4)]-2Q_t \delta Z_V^b,
\nonumber
\end{eqnarray}
\begin{eqnarray}
f_6^{v(\hat{t})}=\frac{1}{2}f_4^{v(\hat{t})}, \nonumber
\end{eqnarray}
\begin{eqnarray}
f_7^{v(\hat{t})}=\frac{\lambda_i^2}{16\pi^2} m_b[
(C_{11}^1+C_{21}^1)-Q_t (C_{11}^2+C_{21}^2)].\nonumber
\end{eqnarray}

\begin{eqnarray}
f_1^{b(\hat{t})}&=&\frac{1}{2} \lambda_i^2 Q_t^2 m_b [ m_t^2
(D_0^1+D_{11}^1)+m_b^2(-D_0^1-2D_{11}^1+D_{12}^1-D_{13}^1-2D_{21}^1
-D_{23}^1+2D_{24}^1
\nonumber\\
&&-2D_{25}^1-D_{31}^1+D_{34}^1-D_{35}^1)+ \hat{s}
(D_{25}^1-D_{26}^1+D_{35}^1-D_{310}^1) + \hat{t} (-D_{11}^1-D_{12}^1
\nonumber\\
&&+D_{13}^1-D_{21}^1-2D_{24}^1+2D_{25}^1-D_{34}^1-D_{35}^1)-4(D_{27}^1+D_{311}^1)]+\lambda_i^2
Q_t m_b(D_{27}^2\nonumber\\
&&+D_{311}^2-D_{312}^2+D_{313}^2)-\lambda_i^2 m_b
D_{311}^3,\nonumber
\end{eqnarray}

\begin{eqnarray}
f_2^{b(\hat{t})}= \lambda_i^2 Q_t^2
m_b(D_{27}^1+D_{311}^1)+\lambda_i^2 Q_t
m_b(D_{27}^2+D_{311}^2-D_{312}^2+D_{313}^2)-\lambda_i^2 m_b
D_{311}^3, \nonumber
\end{eqnarray}
\begin{eqnarray}
f_3^{b(\hat{t})}&=& \lambda_i^2 Q_t^2 [ m_t^2
(-D_{12}^1+D_{13}^1)+m_b^2(D_{12}^1-D_{13}^1-D_{22}^1+3D_{23}^1+2D_{24}^1
-D_{25}^1-D_{26}^1+D_{34}^1
\nonumber\\
&&-D_{35}^1-D_{36}^1-D_{38}^1)+ \hat{s}
(D_{37}^1+D_{38}^1-D_{39}^1-D_{310}^1) + \hat{t}
(D_{22}^1-3D_{23}^1-D_{25}^1+D_{26}^1+D_{36}^1
\nonumber\\
&&+D_{38}^1)+2(D_{27}^1+2D_{312}^1-3D_{313}^1)]+\lambda_i^2 Q_t
[m_t^2 (-D_{11}^2+D_{12}^2)+m_b^2(D_{11}^2-D_{12}^2+D_{21}^2
\nonumber\\
&&-3D_{22}^2-D_{24}^2+2D_{25}^2-2D_{26}^2-D_{32}^2
-D_{34}^2+2D_{35}^2+4D_{36}^2+3D_{37}^2 +2D_{38}^2-5D_{310}^2)
\nonumber\\
&&+ \hat{s}
(2D_{22}^2-D_{36}^2-D_{37}^2-D_{38}^2-D_{39}^2+D_{310}^2)+\hat{t}(D_{21}^2
+D_{22}^2-2D_{24}^2+D_{31}^2-2D_{34}^2
\nonumber\\
&&-D_{36}^2
-D_{37}^2-D_{38}^2-D_{39}^2+D_{310}^2)+2(D_{27}^2+2D_{311}^2-2D_{312}^2)]-2
\lambda_i^2 (D_{27}^3-D_{312}^3),\nonumber
\end{eqnarray}
\begin{eqnarray}
f_4^{b(\hat{t})}&=& \lambda_i^2 Q_t^2 [- m_t^2
D_{13}^1+m_b^2(-D_{13}^1+D_{23}^1-D_{25}^1+D_{35}^1+D_{38}^1-D_{310}^1)+
\hat{s} (-D_{38}^1+D_{39}^1)
\nonumber\\
&&+\hat{t}
(-D_{23}^1+D_{25}^1-D_{38}^1+D_{310}^1)+4D_{311}^1]+2\lambda_i^2 Q_t
(D_{312}^2-D_{313}^2)+2 \lambda_i^2 D_{313}^3,\nonumber
\end{eqnarray}
\begin{eqnarray}
f_5^{b(\hat{t})}&=& \lambda_i^2 Q_t^2 [ m_t^2
(D_{11}^1-D_{12}^1)+m_b^2(-D_{11}^1+D_{12}^1-2D_{21}^1-D_{22}^1+3D_{24}^1
-2D_{26}^1-D_{31}^1+2D_{34}^1
\nonumber\\
&&-D_{35}^1-D_{36}^1-2D_{38}^1-D_{310}^1)+ \hat{s}
(D_{35}^1+D_{37}^1-2D_{310}^1) + \hat{t}
(D_{22}^1-D_{24}^1+2D_{26}^1-D_{34}^1
\nonumber\\
&&+D_{35}^1+D_{36}^1+D_{310}^1)-4(D_{311}^1+D_{312}^1)]+\lambda_i^2
Q_t [m_t^2 D_{13}^2 +m_b^2(-2D_{23}^2-D_{25}^2+D_{26}^2
\nonumber\\
&&-2D_{33}^2-D_{37}^2-2D_{38}^2+3D_{39}^2+D_{310}^2)+ \hat{s}
(D_{33}^2-D_{39}^2)+\hat{t}(-D_{25}^2+D_{26}^2+D_{33}^2-D_{35}^2
\nonumber\\
&&-D_{39}^2+D_{310}^2)-4D_{313}^2] -2\lambda_i^2
(2D_{27}^3+D_{311}^3-D_{312}^3),\nonumber
\end{eqnarray}
\begin{eqnarray}
f_6^{b(\hat{t})}& =& \lambda_i^2 Q_t^2 [m_t^2
D_0^1+m_b^2(D_0^1+2D_{11}^1-2D_{13}^1+D_{21}^1-3D_{25}^1
+D_{26}^1+2D_{38}^1) + \hat{s} (D_{25}^1-D_{26}^1)
\nonumber\\
&&+\hat{t} (D_{25}^1-D_{26}^1)+2(D_{311}^1-D_{313}^1)]+2\lambda_i^2
Q_t (D_{27}^2+D_{311}^2)- 2 \lambda_i^2
(D_{27}^3+D_{311}^3-D_{313}^3),\nonumber
\end{eqnarray}
\begin{eqnarray}
f_7^{b(\hat{t})} &=& 2\lambda_i^2 Q_t^2 m_b(D_{26}^1+D_{310}^1)+2
\lambda_i^2 Q_t
m_b(D_{22}^2-D_{24}^2+D_{25}^2-D_{26}^2-D_{32}^2-D_{34}^2+D_{35}^2
\nonumber\\
&&+2D_{36}^2+ 2D_{37}^2+D_{38}^2-D_{39}^2-3D_{310}^2) +2\lambda_i^2
m_b(D_{25}^3-D_{310}^3),\nonumber
\end{eqnarray}
\begin{eqnarray}
f_8^{b(\hat{t})} &=& 2\lambda_i^2 Q_t^2
m_b(-D_{25}^1+D_{26}^1-D_{35}^1+D_{310}^1) +2\lambda_i^2 Q_t
m_b(-D_{23}^2+D_{26}^2-D_{33}^2-D_{37}^2-D_{38}^2
\nonumber\\
&&+2D_{39}^2+D_{310}^2) + 2 \lambda_i^2
m_b(2D_{25}^3+D_{35}^3-D_{310}^3),\nonumber
\end{eqnarray}
\begin{eqnarray}
f_9^{b(\hat{t})} &=& 2 \lambda_i^2 Q_t^2
m_b(-D_{12}^1-2D_{24}^1+D_{26}^1-D_{34}^1+D_{310}^1)+2 \lambda_i^2
Q_t m_b(-D_{11}^2+D_{12}^2-2D_{21}^2-D_{22}^2
\nonumber\\
&&+3D_{24}^2
-D_{25}^2+D_{26}^2-D_{31}^2+2D_{34}^2-D_{35}^2-D_{36}^2+D_{310}^2)
+2\lambda_i^2 m_b(-D_{11}^3-D_{21}^3+D_{24}^3\nonumber\\
&&+D_{25}^3 +D_{34}^3-D_{310}^3),\nonumber
\end{eqnarray}
\begin{eqnarray}
f_{10}^{b(\hat{t})} &=& 2\lambda_i^2 Q_t^2
m_b(D_{11}^1-D_{12}^1+2D_{21}^1-2D_{24}^1-D_{25}^1+D_{26}^1+D_{31}^1
-D_{34}^1-D_{35}^1+D_{310}^1)
\nonumber\\
&&+ 2\lambda_i^2 Q_t
m_b(D_{23}^2+2D_{25}^2-D_{26}^2+D_{35}^2+D_{38}^2-D_{310}^2) +2
\lambda_i^2 m_b
(-2D_{11}^3-3D_{21}^3+D_{24}^3 \nonumber\\
&&+2D_{25}^3-D_{31}^3+D_{34}^3+D_{35}^3-D_{310}^3),\nonumber
\end{eqnarray}
\begin{eqnarray}
f_{11}^{b(\hat{t})} &=& \frac{1}{2} \lambda_i^2 Q_t^2 [m_t^2
(D_0^1-D_{12}^1+2D_{13}^1)+m_b^2(D_0^1+D_{12}^1-D_{13}^1-D_{21}^1-D_{22}^1
-D_{23}^1+4D_{24}^1
\nonumber\\
&&-2D_{25}^1+D_{31}^1-D_{35}^1-D_{36}^1-D_{38}^1-D_{310}^1)+ \hat{s}
(D_{25}^1-D_{26}^1+D_{37}^1+D_{38}^1+D_{39}^1-D_{310}^1)
\nonumber\\
&&+ \hat{t}
(D_{22}^1+D_{23}^1+D_{36}^1+D_{38}^1-D_{310}^1)+2(2D_{312}^1-D_{313}^1)]-\lambda_i^2
Q_t (D_{311}^2-D_{313}^2)\nonumber\\ &&+ \lambda_i^2
(D_{312}^3-D_{313}^3),\nonumber
\end{eqnarray}
\begin{eqnarray}
f_{12}^{b(\hat{t})}= -\lambda_i^2 Q_t^2
(D_{27}^1+D_{313}^1)-\lambda_i^2 Q_t
(D_{27}^2+D_{311}^2-D_{313}^2)+\lambda_i^2
(D_{312}^3-D_{313}^3),\nonumber
\end{eqnarray}
\begin{eqnarray}
f_{13}^{b(\hat{t})} &=& \lambda_i^2 Q_t^2
m_b(-D_{12}^1+D_{22}^1-D_{24}^1-2D_{25}^1+2D_{26}^1+2D_{310}^1)+\lambda_i^2
Q_t  m_b(-D_{11}^2+D_{12}^2-D_{21}^2
\nonumber\\
&&-D_{22}^2+2D_{24}^2-D_{25}^2+D_{26}^2),\nonumber
\end{eqnarray}
\begin{eqnarray}
f_{14}^{b(\hat{t})}= \lambda_i^2 Q_t^2
m_b(D_{23}^1+D_{25}^1),\nonumber
\end{eqnarray}
\begin{eqnarray}
f_{15}^{b(\hat{t})} = \lambda_i^2 Q_t^2 m_b
(D_{11}^1-D_{12}^1+D_{21}^1+D_{22}^1-D_{24}^1+2D_{26}^1+2D_{310}^1)+\lambda_i^2
Q_t m_b(D_{13}^2+D_{25}^2-D_{26}^2),\nonumber
\end{eqnarray}
\begin{eqnarray}
f_{16}^{b(\hat{t})}= -\lambda_i^2 Q_t^2 m_b
(D_{11}^1-D_{13}^1+D_{21}^1-D_{25}^1),\nonumber
\end{eqnarray}
\begin{eqnarray}
f_{17}^{b(\hat{t})} &=& 2\lambda_i^2 Q_t^2
(D_{23}^1-D_{26}^1-D_{37}^1+D_{39}^1)+ 2\lambda_i^2 Q_t
(-D_{22}^2+D_{24}^2-D_{25}^2+D_{26}^2+D_{34}^2-D_{35}^2
\nonumber\\
&&-D_{36}^2+D_{37}^2+D_{38}^2-D_{39}^2) + 2\lambda_i^2
(D_{23}^3-D_{26}^3+D_{37}^3-D_{39}^3),\nonumber
\end{eqnarray}
\begin{eqnarray}
f_{18}^{b(\hat{t})} &=& 2\lambda_i^2 Q_t^2
(D_{25}^1-D_{26}^1-D_{37}^1-D_{38}^1+D_{39}^1+D_{310}^1)+ 2
\lambda_i^2 Q_t (D_{23}^2-D_{26}^2-D_{33}^2+D_{38}^2+D_{39}^2
\nonumber\\
&&-D_{310}^2) + 2 \lambda_i^2
(2D_{23}^3-2D_{26}^3+D_{37}^3+D_{38}^3-D_{39}^3-D_{310}^3),\nonumber
\end{eqnarray}
\begin{eqnarray}
f_{19}^{b(\hat{t})} &=& 2\lambda_i^2 Q_t^2
(D_{22}^1+D_{23}^1-D_{25}^1+D_{26}^1+D_{36}^1-D_{37}^1+D_{39}^1
+D_{310}^1)+ 2 \lambda_i^2 Q_t (D_{21}^2-D_{25}^2+D_{26}^2
\nonumber\\
&&+D_{31}^2-D_{34}^2-D_{35}^2+D_{310}^2) + 2 \lambda_i^2
(D_{12}^3-D_{13}^3-D_{22}^3+D_{23}^3+D_{24}^3-D_{25}^3
-D_{36}^3+D_{37}^3\nonumber\\&&-D_{39}^3+D_{310}^3),\nonumber
\end{eqnarray}
\begin{eqnarray}
f_{20}^{b(\hat{t})} &=& 2 \lambda_i^2 Q_t^2
(D_{22}^1-D_{24}^1+D_{25}^1+D_{26}^1-D_{34}^1+D_{35}^1+D_{36}^1-D_{37}^1-D_{38}^1+D_{39}^1
+2D_{310}^1)
\nonumber\\
&&+2\lambda_i^2 Q_t  (D_{23}^2-D_{24}^2-D_{25}^2-D_{35}^2+D_{38}^2)
+2 \lambda_i^2 (2D_{12}^3-2D_{13}^3-D_{22}^3+2D_{23}^3+3D_{24}^3
\nonumber\\
&&-3D_{25}^3-D_{26}^3+D_{34}^3-D_{35}^3-D_{36}^3+D_{37}^3+D_{38}^3-D_{39}^3),\nonumber
\end{eqnarray}

\begin{eqnarray}
f_1^{\rm tr}= -\frac{1}{2}\lambda_i^2 m_b C_{11}^5,\nonumber
\end{eqnarray}
where
\begin{eqnarray}
C^1=C(-p_2, p_4, m_t, m_i, m_i), \ \ C^2=C(p_2, -p_4, m_i, m_t,
m_t),\nonumber
\end{eqnarray}
\begin{eqnarray}
C^3=C(p_1, -p_3, m_t, m_i, m_i),\ \ C^4=C(-p_1, p_3, m_i, m_t,
m_t),\nonumber
\end{eqnarray}
\begin{eqnarray}
C^5=C(-p_2, p_1+p_2, m_t, m_i, m_i), \ \ D^1=D(p_2, -p_4, -p_3,
m_i,\nonumber m_t, m_t, m_t),
\end{eqnarray}
\begin{eqnarray}
D^2=D(-p_1+p_3, p_1, -p_1+p_4, m_i, m_t, m_i, m_t),\ \ D^3=D(-p_2,
p_4, p_3, m_t, m_i, m_i, m_i),\nonumber
\end{eqnarray}
and
\begin{eqnarray}
\hat{s}=(p_1+p_2)^2,\ \ \hat{t}=(p_4-p_2)^2,\ \
 \hat{u}=(p_4-p_1)^2. \nonumber
\end{eqnarray}
For $i=\pi$,
\begin{eqnarray}
\lambda_{\pi}=\frac{c_t \varepsilon m_t}{f_{\pi}},\nonumber
\end{eqnarray}
for $i=\pi_8$,
\begin{eqnarray}
\lambda _{\pi_8}=\frac{\varepsilon m_t \lambda^a}{f_{\pi}},\nonumber
\end{eqnarray}
and for $i=\pi_t$,
\begin{eqnarray}
\lambda_t=\frac{(1-\varepsilon) m_t}{f_{\pi_t}}.\nonumber
\end{eqnarray}

\section*{APPENDIX B: THE FORM FACTORS ARISING FROM NEW GAUGE BOSONS}

\ \ \ The form factors $f_i^{s(\hat{t})}$ from new gauge bosons
($Z^*$, $Z'$, and $B$) can be written as
\begin{eqnarray}
f_1^{s(\hat{t})}=-2m_b(p_2 \cdot p_4)
[\Sigma_S^b(\hat{t})-\frac{\delta m_b}{m_b}-\delta Z_V^b]-2m_b(p_2
\cdot p_4) [\Sigma_V^b(\hat{t})+\delta Z_V^b],\nonumber
\end{eqnarray}
\begin{eqnarray}
f_2^{s(\hat{t})}=4m_b^2 [\Sigma_S^b(\hat{t})-\frac{\delta
m_b}{m_b}-\delta Z_V^b]+4(m_b^2-p_2 \cdot p_4)
[\Sigma_V^b(\hat{t})+\delta Z_V^b],\nonumber
\end{eqnarray}
\begin{eqnarray}
f_3^{s(\hat{t})}=\frac{1}{2}f_2^{s(\hat{t})},\nonumber
\end{eqnarray}
where
\begin{eqnarray}
\Sigma_V^b(p^2)=-\frac{1}{16\pi^2}(\lambda_1^2+\lambda_2^2)B_1(p^2,
m_b, m_i),\nonumber
\end{eqnarray}
\begin{eqnarray}
\Sigma_S^b(p^2)=-\frac{1}{4\pi^2}\lambda_1 \lambda_2 B_0(p^2, m_b,
m_i),\nonumber
\end{eqnarray}
\begin{eqnarray}
\delta m_b=m_b[\Sigma_V^b(m_b^2)+\Sigma_S^b(m_b^2)],\nonumber
\end{eqnarray}
and
\begin{eqnarray}
\delta Z_V^b=-\Sigma_V^b(m_b^2)-2m_b^2\frac{\partial}{\partial
p^2}[\Sigma_V^{b}(p^2)+\Sigma_S^{b}(p^2)]|_{p^2=m_b^2}.\nonumber
\end{eqnarray}

 The form factors $f_i^{v(\hat{t})}$, and
$f_i^{b(\hat{t})}$ are given by
\begin{eqnarray}
f_1^{v(\hat{t})}=\frac{1}{8\pi^2} m_b p_2 \cdot p_4 Q_b
(\lambda_1^2+\lambda_2^2) (C_0^1+C_{11}^1+C_0^2+C_{11}^2),\nonumber
\end{eqnarray}
\begin{eqnarray}
f_2^{v(\hat{t})}=\frac{1}{4\pi^2} p_2 \cdot p_4 Q_b (\lambda_1^2
+\lambda_2^2) (C_{0}^2+C_{11}^2+C_{12}^2+C_{23}^2)],\nonumber
\end{eqnarray}
\begin{eqnarray}
f_3^{v(\hat{t})}&=&\frac{1}{8\pi^2}Q_b [2 m_b^2
(\lambda_1^2+\lambda_2^2) (C_{0}^1+C_{11}^1-C_{0}^2-2C_{11}^2)+2 p_2
\cdot p_4 (\lambda_1^2+\lambda_2^2)
(C_{0}^2+C_{11}^2+C_{12}^2\nonumber\\&& +C_{23}^2)+m_b^2
(\lambda_1^2+\lambda_2^2) (C_{21}^1-C_{21}^2)- 8 m_b^2 \lambda_1
\lambda_2 (C_0^1+C_{11}^1) -2 (\lambda_1^2+\lambda_2^2)
(C_{24}^1+C_{24}^2)]\nonumber\\&& -4Q_b \delta Z_V^b,\nonumber
\end{eqnarray}
\begin{eqnarray}
f_4^{v(\hat{t})}=\frac{1}{4\pi^2} m_b Q_b[4\lambda_1
\lambda_2(C_0^2+C_{11}^2)
-(\lambda_1^2+\lambda_2^2)(2C_{0}^2+3C_{11}^2+C_{21}^2)],\nonumber
\end{eqnarray}
\begin{eqnarray}
f_5^{v(\hat{t})}&=& \frac{1}{16\pi^2}Q_b [-2m_b^2
(\lambda_1^2+\lambda_2^2)(C_0^1+2C_{11}^1+C_0^2+2C_{11}^2)
-m_b^2(\lambda_1^2+\lambda_2^2)(C_{21}^1+C_{21}^2) \nonumber \\
&&+2 p_2\cdot p_4
(\lambda_1^2+\lambda_2^2)(C_0^1+C_{11}^1+C_{12}^1+C_{23}^1+
C_0^2+C_{11}^2+C_{12}^2+C_{23}^2) \nonumber \\ && -2
(\lambda_1^2+\lambda_2^2)(C_{24}^1+C_{24}^2)]-2Q_b \delta Z_V^b,
\nonumber
\end{eqnarray}
\begin{eqnarray}
f_6^{v(\hat{t})}=\frac{1}{2}f_4^{v(\hat{t})}, \nonumber
\end{eqnarray}
\begin{eqnarray}
f_7^{v(\hat{t})}=\frac{1}{8\pi^2} m_b Q_b[4\lambda_1
\lambda_2(C_0^1+C_{11}^1)
-(\lambda_1^2+\lambda_2^2)(2C_{0}^1+3C_{11}^1+C_{21}^1)].\nonumber
\end{eqnarray}

\begin{eqnarray}
f_1^{b(\hat{t})}&=& Q_b^2 [m_b^3 (\lambda_1^2+\lambda_2^2)
(2D_0+2D_{11}+D_{12}-2D_{13}+2D_{23}-D_{25})-m_b
\hat{s}(\lambda_1^2+\lambda_2^2) (D_0+D_{11}
\nonumber\\
&& -D_{13}+D_{23})-m_b
\hat{t}(\lambda_1^2+\lambda_2^2)(2D_{0}+2D_{11}+D_{12}-2D_{13}+2D_{23}-D_{25})
 \nonumber \\
&&-2 m_b (\lambda_1^2+\lambda_2^2) (D_{27}-D_{313})+2 m_b^3
\lambda_1 \lambda_2 (2 D_{11}-D_{12}+D_{25})-2m_b \hat{s}
\lambda_1\lambda_2 (D_0+D_{25} \nonumber \\&& -D_{26})+ 2m_b \hat{t}
\lambda_1\lambda_2 (2D_{0}+D_{12}+D_{21}+D_{24}-D_{25})] \nonumber,
\end{eqnarray}
\begin{eqnarray}
f_2^{b(\hat{t})}&=& Q_b^2 [-m_b^3 (\lambda_1^2+\lambda_2^2)
(D_{11}-D_{12}+D_{13}+2D_{21}-2D_{23}-2D_{24}+5D_{25}+D_{31}+D_{34}+D_{35}
\nonumber\\
&& -2D_{38}-2D_{39}+2D_{310})-m_b \hat{s}(\lambda_1^2+\lambda_2^2)
(2 D_{23}-3D_{25}+D_{26}-D_{35}+2D_{38}+2D_{39}
\nonumber\\
&& -D_{310})-m_b
\hat{t}(\lambda_1^2+\lambda_2^2)(D_{11}+D_{12}-D_{13}+D_{21}
+2D_{23}+2D_{24}-4D_{25}-D_{34}-D_{35}
 \nonumber \\
&& +2D_{38}+2D_{39}-2D_{310})- 4 m_b (\lambda_1^2+\lambda_2^2)
(D_{27}+D_{311})+2 m_b^2 \lambda_1 \lambda_2 (2 D_0+D_{11}-D_{12}
\nonumber \\&& -3D_{13}+D_{21}-D_{24}+D_{25}) +2m_b \hat{s}
\lambda_1\lambda_2 (2D_{13}-D_{25}+D_{26})+ 2m_b \hat{t}
\lambda_1\lambda_2 (D_{11}+D_{12}\nonumber \\&&
-D_{13}+D_{24}-D_{25})] \nonumber,
\end{eqnarray}
\begin{eqnarray}
f_3^{b(\hat{t})}&=& Q_b^2 [2 m_b^2(\lambda_1^2+\lambda_2^2)
(2D_{11}+2D_{21}-3D_{23}-D_{24}+3D_{25}+D_{35}-D_{38}-D_{310})
\nonumber\\
&&
-2\hat{s}(\lambda_1^2+\lambda_2^2)(D_{13}-D_{23}+D_{24}+2D_{25}-D_{26}-D_{39})
+2\hat{t}(\lambda_1^2+\lambda_2^2)(2D_{12}-2D_{13}
\nonumber\\
&& +D_{24}-D_{25} +D_{38}+D_{310})- 4 (\lambda_1^2+\lambda_2^2)
(D_{27}+D_{312}-3D_{313})] \nonumber,
\end{eqnarray}
\begin{eqnarray}
f_4^{b(\hat{t})}&=& Q_b^2 [-2 m_b^2(\lambda_1^2+\lambda_2^2)
(2D_{11}-D_{12}+D_{21}-D_{23}-D_{24}-3D_{25}+D_{26}) -4
\hat{t}(\lambda_1^2+\lambda_2^2)(D_{12}
\nonumber\\
&& -D_{13}+D_{24}-D_{26})-4 (\lambda_1^2+\lambda_2^2)
(D_{27}+D_{313})+4 m_b^2 \lambda_1\lambda_2 (2D_0+D_{13})]
\nonumber,
\end{eqnarray}
\begin{eqnarray}
f_5^{b(\hat{t})}&=& Q_b^2 [-2 m_b^2(\lambda_1^2+\lambda_2^2)
(2D_0+D_{12}-D_{13}-2D_{21}+D_{22}+D_{24}-2D_{25}-D_{26}) +2
\hat{t}(\lambda_1^2
\nonumber\\
&& +\lambda_2^2)(D_{12}-D_{13}-D_{21}+D_{22}+D_{24}-D_{26})+ 4
(\lambda_1^2+\lambda_2^2) (D_{27}+D_{311}-D_{312})] \nonumber,
\end{eqnarray}
\begin{eqnarray}
f_6^{b(\hat{t})}&=& Q_b^2 [-2 m_b^2(\lambda_1^2+\lambda_2^2)
(2D_{0}+6D_{11}-2D_{12}-5D_{21}-D_{23}-3D_{24}-2D_{25}+2D_{26}+D_{31}
\nonumber\\
&& -D_{34}-D_{38}+D_{310}) +2
\hat{s}(\lambda_1^2+\lambda_2^2)(D_{0}+2D_{11}-D_{12}+D_{21}-D_{24}+D_{35}-D_{38}
\nonumber\\
&& +D_{39}-D_{310})
-2\hat{t}(\lambda_1^2+\lambda_2^2)(2D_{12}-2D_{13}
+D_{23}+3D_{24}-2D_{25}-2D_{26}+D_{34}-D_{35}
\nonumber\\
&& +D_{38}-D_{310})- 8 (\lambda_1^2+\lambda_2^2)
(D_{27}+D_{311}-D_{313})+4 m_b^2 \lambda_1\lambda_2 (2D_0+D_{13})]
\nonumber,
\end{eqnarray}
\begin{eqnarray}
f_7^{b(\hat{t})}= 4 m_b Q_b^2 [(\lambda_1^2+\lambda_2^2)
(2D_{13}+2D_{25}+D_{26}+D_{310})-2\lambda_1\lambda_2
(D_{13}+2D_{26})] \nonumber,
\end{eqnarray}
\begin{eqnarray}
f_8^{b(\hat{t})}= 4 m_b Q_b^2 [(\lambda_1^2+\lambda_2^2)
(D_{26}-D_{35}+D_{310})+4 \lambda_1\lambda_2 (D_{25}-D_{26})]
\nonumber,
\end{eqnarray}
\begin{eqnarray}
f_9^{b(\hat{t})}&=& 4 m_b Q_b^2 [-(\lambda_1^2+\lambda_2^2)
(D_0+3D_{11}+D_{12}-2D_{13}+2D_{21}+2D_{24}-2D_{25}-D_{26}+D_{34}
\nonumber \\ &&-D_{310}) +2 \lambda_1\lambda_2
(2D_0+2D_{11}+2D_{12}-D_{13}+2D_{24}-2D_{26})] \nonumber,
\end{eqnarray}
\begin{eqnarray}
f_{10}^{b(\hat{t})}&=& 4 m_b Q_b^2 [(\lambda_1^2+\lambda_2^2)
(D_0+2D_{11}-D_{12}+2D_{21}-2D_{24}-D_{25}+D_{26}+D_{31}-D_{34}
\nonumber
\\ && -D_{35}+D_{310}) -2 \lambda_1\lambda_2
(D_0+2D_{11}-2D_{12}+2D_{21}-2D_{24}-2D_{25}+2D_{26})] \nonumber,
\end{eqnarray}
\begin{eqnarray}
f_{11}^{b(\hat{t})}&=& Q_b^2 [-4 m_b^2(\lambda_1^2+\lambda_2^2)
(D_0+D_{11}+D_{12})+\hat{s}(\lambda_1^2+\lambda_2^2)(D_{0}+D_{11})
-2(\lambda_1^2+\lambda_2^2) (D_{27}\nonumber
\\ && +D_{312}-D_{313})+2 m_b^2 \lambda_1\lambda_2 D_{13}] \nonumber,
\end{eqnarray}
\begin{eqnarray}
f_{12}^{b(\hat{t})}&=& Q_b^2 [m_b^2(\lambda_1^2+\lambda_2^2)
(2D_0-D_{13}-D_{21}-D_{22}-D_{23}+2D_{24}-2D_{25}+2D_{26}+D_{34}
\nonumber
\\ &&-D_{35}-D_{36}-D_{38}+2D_{310})+\hat{s}(\lambda_1^2
+\lambda_2^2)(D_{25}-D_{26}+D_{37}+D_{38}-D_{39}-D_{310}) \nonumber
\\ && +\hat{t}(\lambda_1^2+\lambda_2^2) (D_{22}+D_{23}-2D_{26}+D_{36}+D_{38} -2D_{310})
+4(\lambda_1^2+\lambda_2^2)(D_{312}-D_{313})\nonumber
\\ && -2 m_b^2
\lambda_1\lambda_2 (2D_0+3D_{13})] \nonumber,
\end{eqnarray}
\begin{eqnarray}
f_{13}^{b(\hat{t})} =  2 m_b Q_b^2 [(\lambda_1^2+\lambda_2^2)
(D_0+D_{11}-D_{12}-D_{24})+2 \lambda_1\lambda_2 (D_{0}+D_{13})]
\nonumber,
\end{eqnarray}
\begin{eqnarray}
f_{14}^{b(\hat{t})} =2 m_b Q_b^2 [(\lambda_1^2+\lambda_2^2)
(D_0+D_{11}-D_{13}-D_{25})-2 \lambda_1\lambda_2 (D_{0}+D_{13})]
\nonumber,
\end{eqnarray}
\begin{eqnarray}
f_{15}^{b(\hat{t})} = 2 m_b Q_b^2 [(\lambda_1^2+\lambda_2^2)
(D_0+D_{12}-D_{21})-2 \lambda_1\lambda_2 D_{0}] \nonumber,
\end{eqnarray}
\begin{eqnarray}
f_{16}^{b(\hat{t})} = 2 m_b Q_b^2 [-(\lambda_1^2+\lambda_2^2)
(D_0+D_{13}-D_{21}+D_{25})+2 \lambda_1\lambda_2 D_{0}] \nonumber,
\end{eqnarray}
\begin{eqnarray}
f_{17}^{b(\hat{t})} = 4 Q_b^2 (\lambda_1^2+\lambda_2^2)
(D_{23}-D_{26}-D_{37}+D_{39}) \nonumber,
\end{eqnarray}
\begin{eqnarray}
f_{18}^{b(\hat{t})} =- 4 Q_b^2 (\lambda_1^2+\lambda_2^2)
(D_{25}-D_{26}+D_{37}+D_{38}-D_{39}-D_{310}) \nonumber,
\end{eqnarray}
\begin{eqnarray}
f_{19}^{b(\hat{t})} &=& 4 Q_b^2 (\lambda_1^2+\lambda_2^2)
(2D_{12}-2D_{13}+D_{22}+D_{23}+2D_{24}-D_{25}-3D_{26}+D_{36}-D_{37}
\nonumber \\ && +D_{39}-D_{310}) \nonumber,
\end{eqnarray}
\begin{eqnarray}
f_{20}^{b(\hat{t})} = 4 Q_b^2 (\lambda_1^2+\lambda_2^2)
(D_{22}-D_{24}-D_{34}+D_{35}+D_{36}-D_{37}-D_{38}+D_{39}) \nonumber,
\end{eqnarray}
with
\begin{eqnarray}
 C^1=C(p_2, -p_4, m_i, m_b,
m_b),\ \ C^2=C(-p_1, p_3, m_i, m_b, m_b),\nonumber
\end{eqnarray}
\begin{eqnarray}
D=D(p_2, -p_4, -p_3, m_i, m_b, m_b, m_b).\nonumber
\end{eqnarray}

For $i=Z^*$,
\begin{eqnarray}
\lambda_1=-\frac{\varepsilon m_t}{16\pi f_{\pi}} \frac{e}{s_Wc_W}
[\frac{N_C}{N_{TC}+1}\xi_t(\xi_t^{-1}+\xi_b)-\xi_t^2],\ \
\lambda_2=0,\nonumber
\end{eqnarray}
for $i=Z'$,
\begin{eqnarray}
\lambda_1=\frac{1}{6}g_1 \cot \theta',\ \ \lambda_2=- \frac{1}{3}g_1
\cot \theta', \nonumber
\end{eqnarray}
and for $i=B$,
\begin{eqnarray}
\lambda_1=\lambda_2= \frac{1}{2} g_3 \cot\theta \lambda^a. \nonumber
\end{eqnarray}


\newpage

\section*{Figure captions}

\ \ \ Fig.1 \ Feynman diagrams for PGB contributions to the
$\gamma\gamma \rightarrow b\bar{b}$ process: (a)-(b) tree-level
diagrams; (c)-(e) self-energy diagrams; (f)-(i) vertex diagrams;
(j)-(l) box diagrams; (m) triangle diagram. Here only one-loop
diagrams corresponding to the tree-level diagram (a) are plotted.
The dashed lines represent the charged technipions $\pi^{\pm}$,
$\pi_8^{\pm}$ and top pions $\pi_t^{\pm}$ in the figures (c)-(m).

\

Fig.2 \ Feynman diagrams for the contributions arising from new
gauge bosons to the $\gamma\gamma \rightarrow b\bar{b}$ process:
(a)-(c) self-energy diagrams; (d)-(e) vertex diagrams; (f) box
diagram. Here only one-loop diagrams corresponding to the tree-level
$t-$channel diagram are plotted. The folding lines denote the new
gauge bosons ($Z^*$, $Z'$ and $B$).

\

Fig.3 \ The relative correction $\delta \sigma (e^+e^- \rightarrow
\gamma\gamma \rightarrow b\bar{b})$ curves as a function of
$\varepsilon$ for $m_{\pi} = 150$ GeV, $m_{\pi_8} = 246$ GeV, and
$m_{\pi_8} = 246$ GeV.

\

Fig.4 \ The relative correction $\delta \sigma (e^+e^- \rightarrow
\gamma\gamma \rightarrow b\bar{b})$ vs $m_{\pi_t}$, when
$\varepsilon=0.06$, $m_{\pi}=150$ GeV, and $m_{\pi_8}=246$ GeV.

\

Fig.5 \ The total cross sections $\sigma(s)$ arising from PGBs in
the TOPCTC model as a function of $\sqrt{s}$ with
$\varepsilon=0.06$, $m_{\pi}=150$ GeV,  $m_{\pi_8}=246$ GeV, and
$m_{\pi_t}=225$ GeV.

\

Fig.6 \ The relative correction $\delta \sigma (e^+e^- \rightarrow
\gamma\gamma \rightarrow b\bar{b})$ vs $m_{Z'}$ when $\kappa_1=1$,
$4$ and $8$.

\ \ \ \

\newpage

\

 \begin{figure}
 \begin{center}
 \begin{picture}(300,20)(0,300)
 \put(-100,+50){\epsfxsize170mm\epsfbox{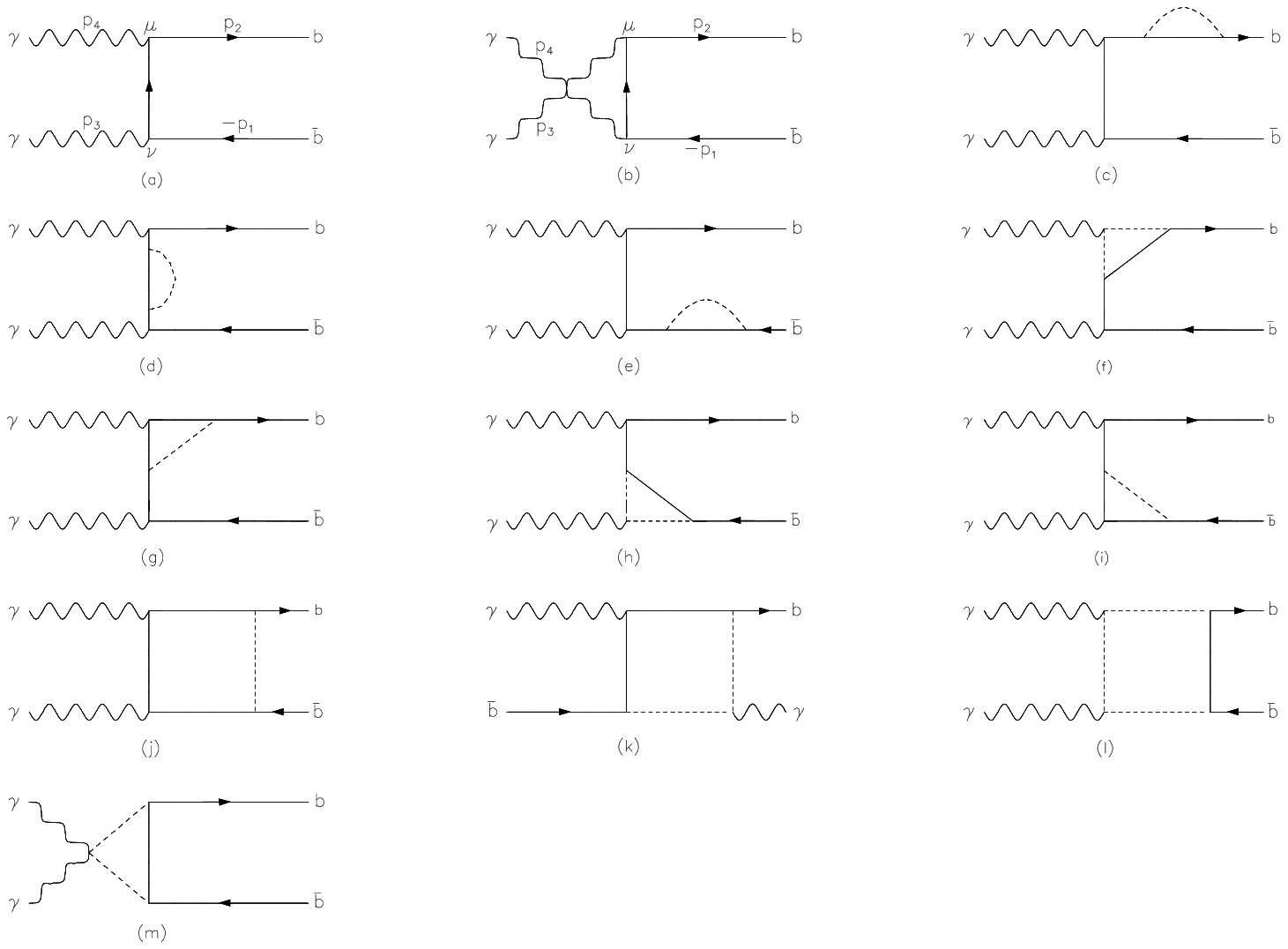}}
\put(140,+30){FIG. 1}
 \put(-100,-330){\epsfxsize170mm\epsfbox{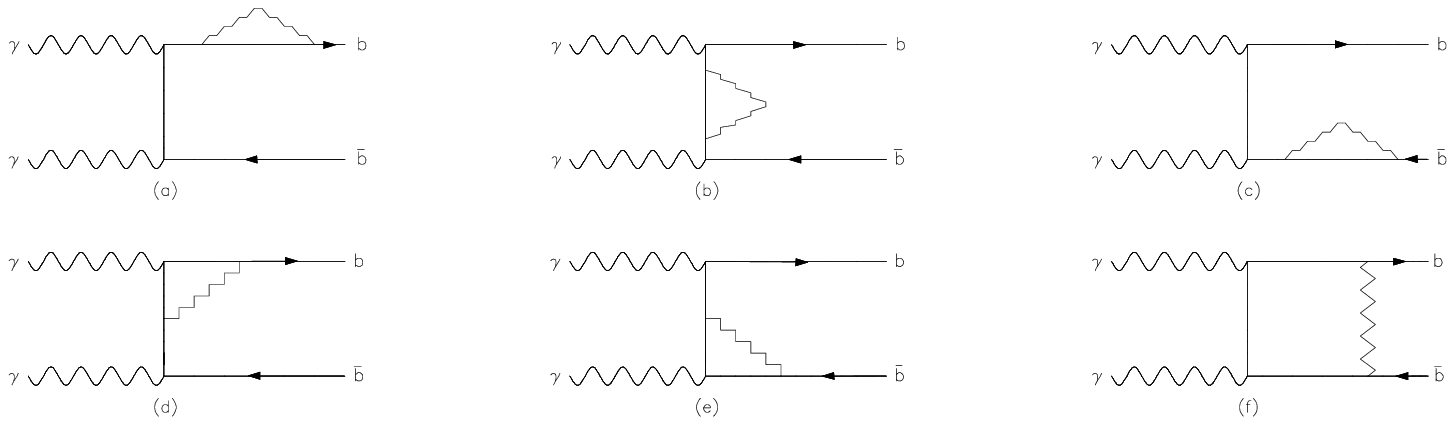}}
\put(140,-240){FIG. 2}
 \end{picture}
 \end{center}
 \end{figure}

\newpage

\

 \begin{figure}
 \begin{center}
 \begin{picture}(300,20)(0,300)
 \put(-80,+60){\epsfxsize150mm\epsfbox{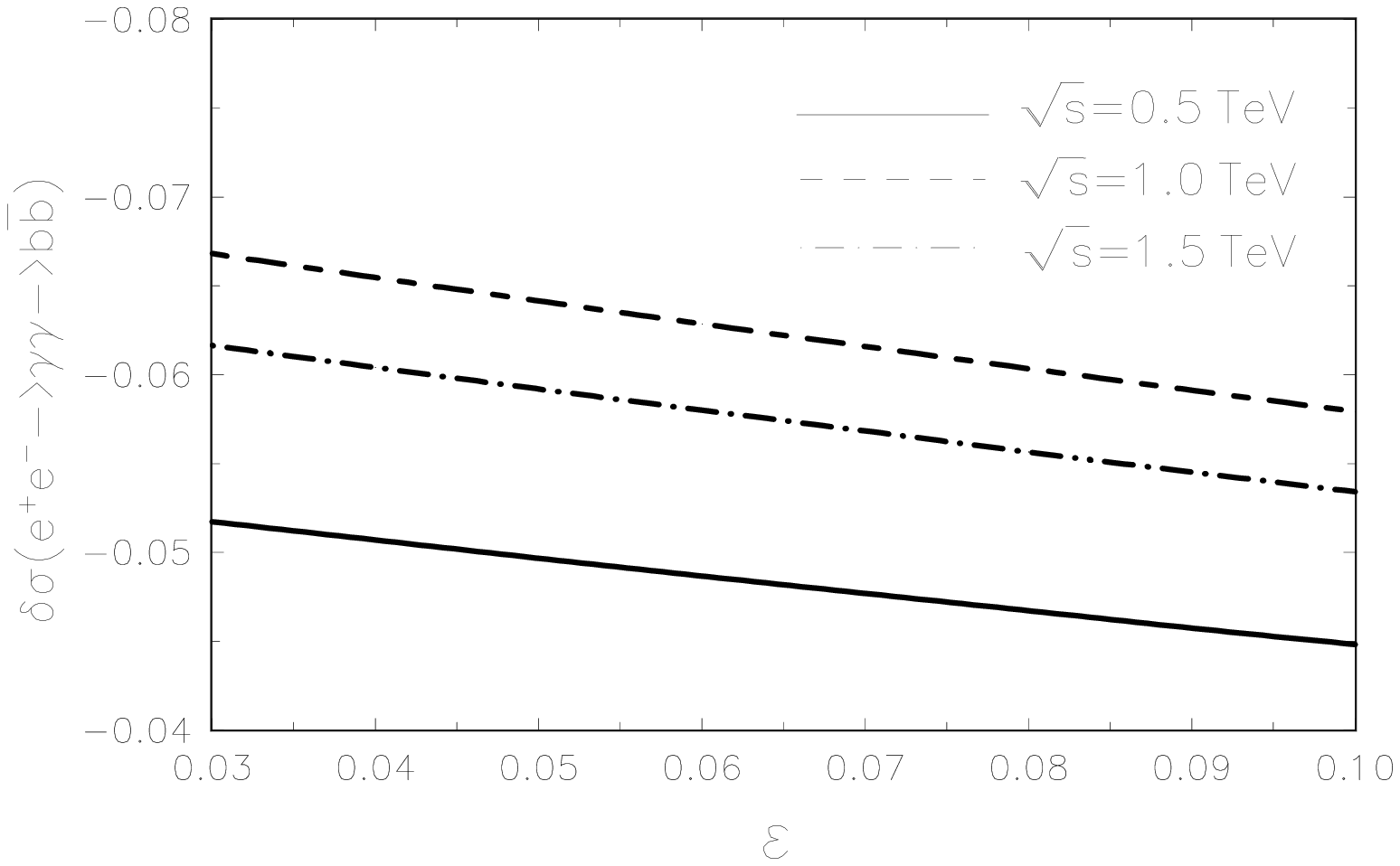}}
\put(140,+40){FIG. 3}
 \put(-80,-260){\epsfxsize150mm\epsfbox{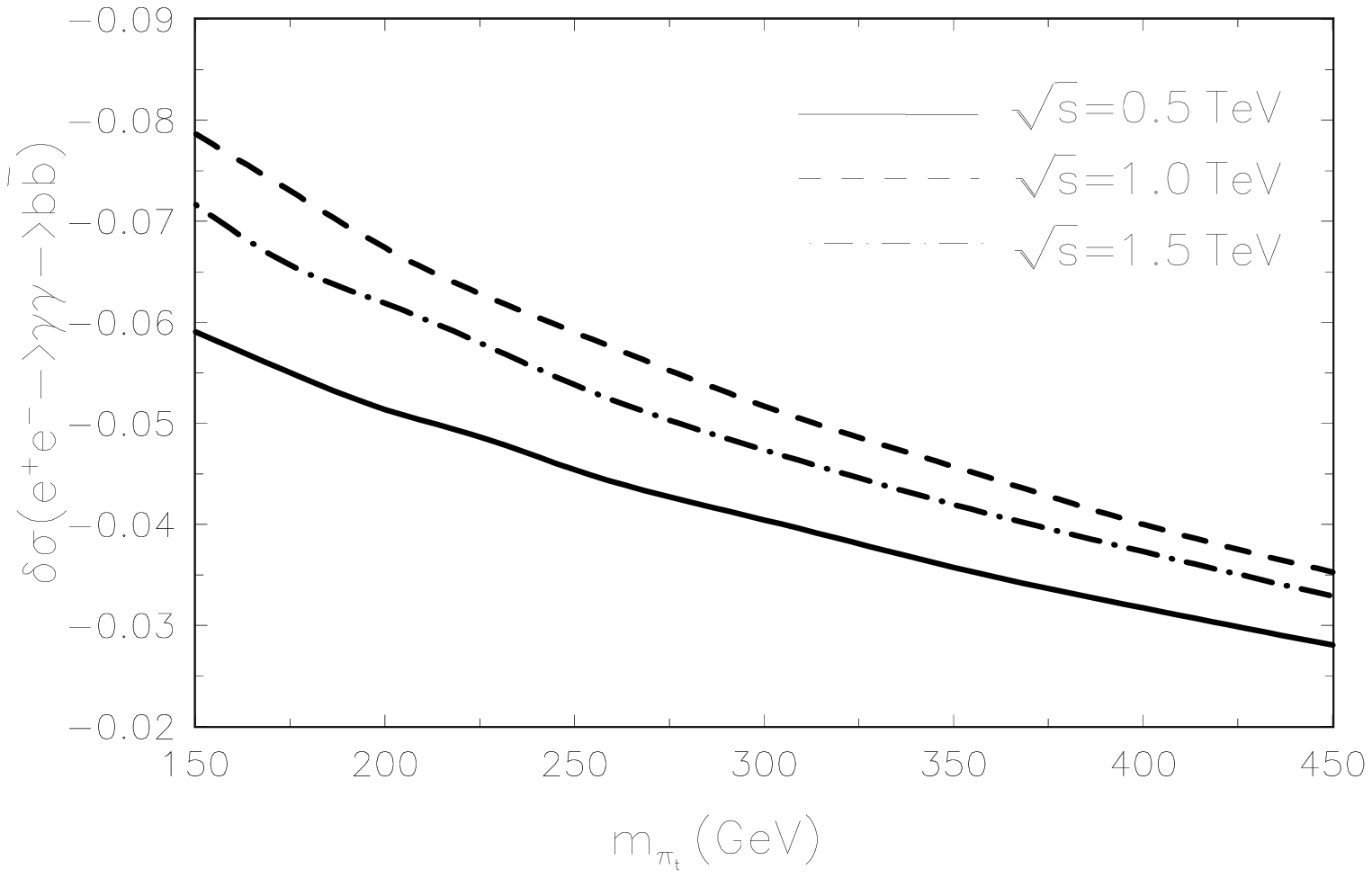}}
\put(140,-280){FIG. 4}
  \end{picture}
 \end{center}
 \end{figure}

\newpage

\

 \begin{figure}
 \begin{center}
 \begin{picture}(300,20)(0,300)
 \put(-80,+60){\epsfxsize150mm\epsfbox{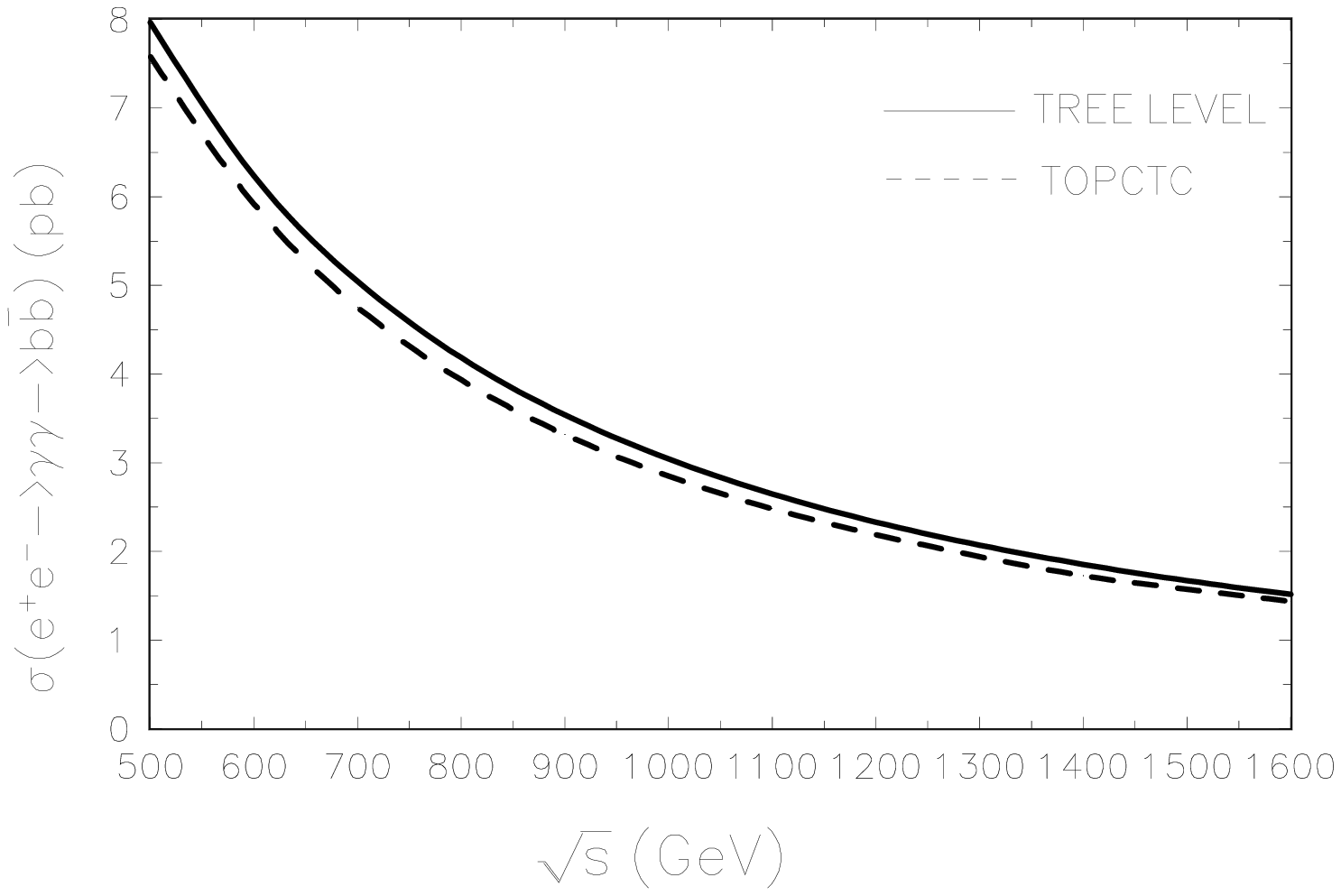}}
\put(140,+40){FIG. 5}
 \put(-80,-260){\epsfxsize150mm\epsfbox{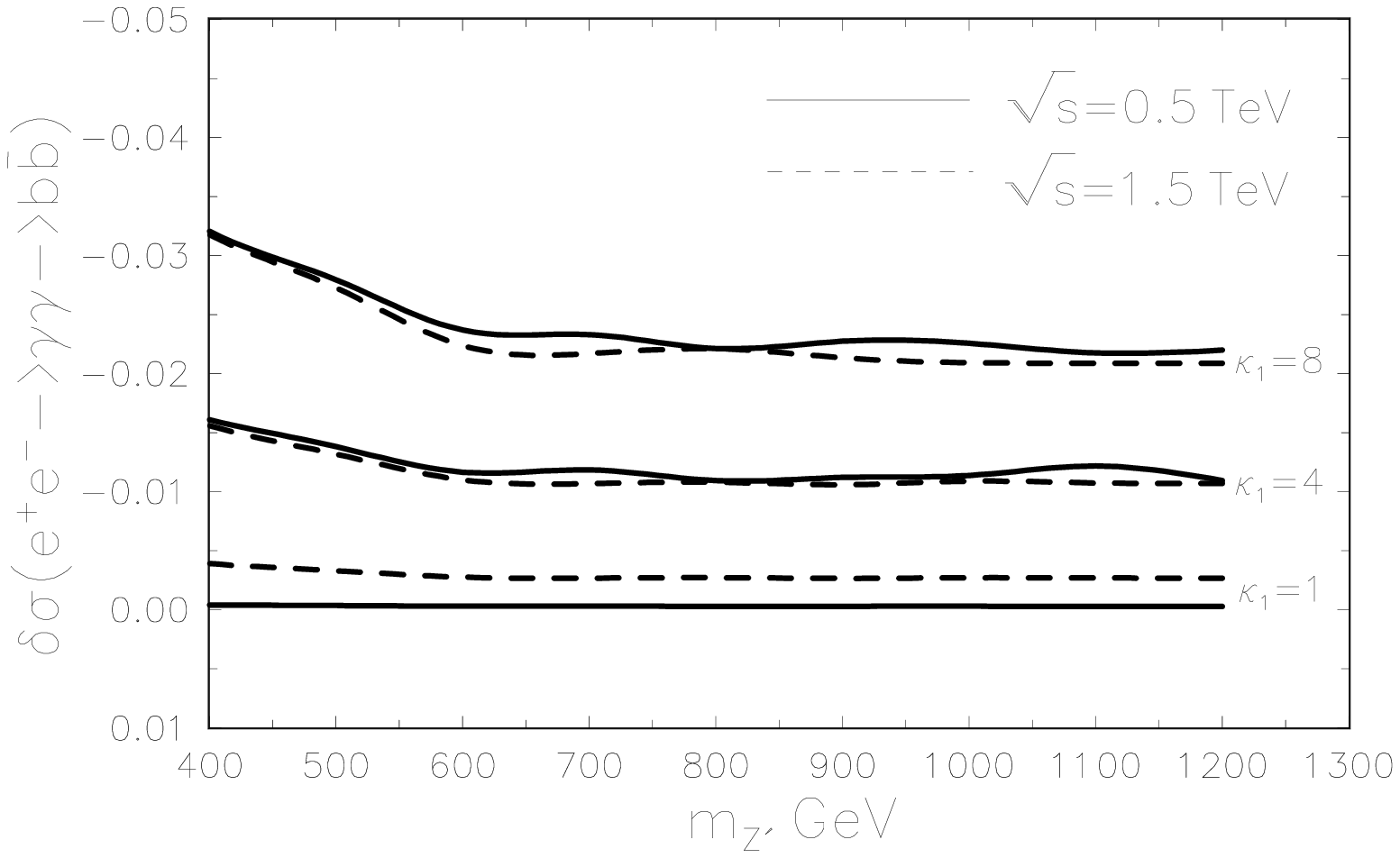}}
\put(140,-280){FIG. 6}
  \end{picture}
 \end{center}
 \end{figure}

\end{document}